\documentclass[conference, anonymous]{IEEEtran}
\IEEEoverridecommandlockouts
\usepackage{cite}
\usepackage{amsmath,amssymb,amsfonts}
\usepackage{algorithmic}
\usepackage{graphicx}
\usepackage{textcomp}
\usepackage{xcolor}
\usepackage{subcaption}
\usepackage{booktabs, multirow} 
\usepackage{soul}
\usepackage{xcolor,colortbl} 
\usepackage{changepage,threeparttable} 

\def\BibTeX{{\rm B\kern-.05em{\sc i\kern-.025em b}\kern-.08em
    T\kern-.1667em\lower.7ex\hbox{E}\kern-.125emX}}

\newcommand{\sys}{\textit{MergeGuard}}

\begin{document}

\title{MergeGuard: Efficient Thwarting of Trojan Attacks in Machine Learning Models
}

\author{
    \IEEEauthorblockN{Soheil Zibakhsh Shabgahi\textsuperscript{*}, Yaman Jandali\textsuperscript{*}, Farinaz Koushanfar} 
    \IEEEauthorblockA{\textit{Department of Electrical and Computer Engineering} \\
    \textit{University of California San Diego}\\
    La Jolla, CA, USA \\
    \{szibakhshshabgahi, yeljanda, fkoushanfar\}@ucsd.edu}
}


\maketitle
\begingroup
\renewcommand\thefootnote{}\footnotetext{*Equal contribution.}
\endgroup

\begin{abstract}
This paper proposes \sys, a novel methodology for mitigation of AI Trojan attacks. Trojan attacks on AI models cause inputs embedded with triggers to be misclassified to an adversary’s target class, posing a significant threat to model usability trained by an untrusted third party. The core of \sys{} is a new post-training methodology for linearizing and merging fully connected layers which we show simultaneously improves model generalizability and performance. Our Proof of Concept evaluation on Transformer models demonstrates that \sys{} maintains model accuracy while decreasing trojan attack success rate, outperforming commonly used (post-training) Trojan mitigation by fine-tuning methodologies. 
\end{abstract}


\section{Introduction}
\label{sec:introduction}

Utilizing Artificial Intelligence (AI) for automation is increasingly ingrained in various technical fields. Recent research has shown that larger Deep Neural Networks (DNNs) with greater expressive capacity can more effectively approximate complex real-world functions and achieve higher accuracy \cite{modelSize1, modelSize2}. As model architectures grow in size, so too do the datasets required to train these data-hungry models. To conserve resources, modern Machine Learning (ML) practitioners frequently rely on pretrained models or publicly available datasets, exposing themselves to the risk of maliciously manipulated models or tampered datasets. Numerous studies \cite{trojan1, trojan2} have demonstrated the feasibility of training \textit{trojaned models}, where an adversary injects triggers into the model to force specific inputs to be misclassified to a target label. 

With the increasing deployment of ML models in critical applications such as in autonomous vehicles, medical diagnostics, and financial decision-making, \cite{autoDriving, medical, finance} the potential consequences of such attacks continue to become more severe. 

This growing concern regarding the safe deployment of AI models has prompted a surge of research focused on trojan detection and mitigation \cite{trojan2, trojan3, trojan4, trojan5}. A straightforward approach involves using standard fine-tuning, which may be effective for simpler models or against more subtle attacks. However, Wu et al. \cite{backdoorbench} found that vanilla fine-tuning often struggles as a mitigation strategy when applied using smaller benign datasets against more sophisticated, powerful attacks. As a result, more advanced mitigation strategies have been proposed, typically leveraging computationally heavy techniques or enhancements of fine-tuning tailored for trojan removal. While many of these methods have shown promise in mitigating trojans for smaller models, such as PreAct-ResNet18 \cite{ftsam, neuralcleanse}, there remains a significant gap in research addressing the mitigation of trojans in larger transformer based architectures, such as Vision Transformers.

The Vision Transformer (ViT) is growing in popularity due to its exeptional performance and ability to extend beyond vision to audio, video, and other applications \cite{audioViT, videoViT}. Subramanya et al. \cite{visionTransformer} observed that vision transformer architectures are especially vulnerable to trojan attacks, a finding corroborated by our own experiments. Given the substantial size of transformer-based models and the extensive datasets required for their training, these models and their training datasets are frequently obtained from online sources. This reliance on external pretrained models and datasets makes them especially prone to manipulation. Therefore, it is crucial to develop effective mitigation strategies that reliably generalize to larger architectures, such as those of transformer based models, to ensure the safe deployment of ML models in real-world applications. While prior state-of-the-art methods have shown promise in detecting and mitigating trojans in smaller convolutional models, they often struggle when extended to transformer architectures. 


In this work, we propose a novel architecture-agnostic mitigation strategy, \sys, which we demonstrate to be both effective and computationally efficient for vision transformers as well as CNN architectures. Our approach not only shows enhanced resilience against trojan attacks but also preserves the high performance of the underlying model, making it suitable for deployment in high-stakes environments. Additionally, we show that \sys{} serves a dual purpose of post-training compression, achieving up to a 15\% reduction in model size for transformers without compromising accuracy. 
\sys{} acts as an effective trojan mitigation strategy by means of post-training compression, removing the layers containing backdoor-related neurons. A regularization term controls the rate of compression, optimizing for model size and performance simultaneously. This regularizer reduces the model depth by adaptively linearizing the activation functions between fully connected layers, allowing for the fusion of adjacent linear layers. This effectively reduces the overall depth and complexity of the model without significantly impacting its performance.

In summary, our contributions are as follows:
\begin{itemize}
    \item We demonstrate the poor generalizability of a number of works targeted at cleansing CNN-based models applied to trojaned transformer models.
    \item We introduce \sys{}, a novel, model-agnostic trojan mitigation method relying on post-processing and showcase its robust performance in mitigating a range of trojan attacks.
    \item We evaluate the effectiveness of \sys{} across different architectures, including CNNs, establishing its versatility beyond transformer models. 
    \item We show that \sys{} achieves significant computational efficiency, offering speedups up to 17.7x compared to other methods, making it a practical solution for large-scale deployment.
    \item We demonstrate \sys{}'s capability as a compression-aware regularizer, achieving up to $15\%$ parameter reduction and $14\%$ MAC reduction in transformer models without compromising accuracy.
    \item Our code is accessible to support reproducibility and further advancements in trojan mitigation research.\footnote{https://github.com/yjandali/BackdoorBench-MergeGuard}
\end{itemize}

\section{Background and Related Work}
\label{sec:relatedwork}

\subsection{Trojan Attacks}

\begin{figure}[h]
    \centering
    \begin{minipage}{0.45\linewidth}
        \centering
        \includegraphics[width=\linewidth]{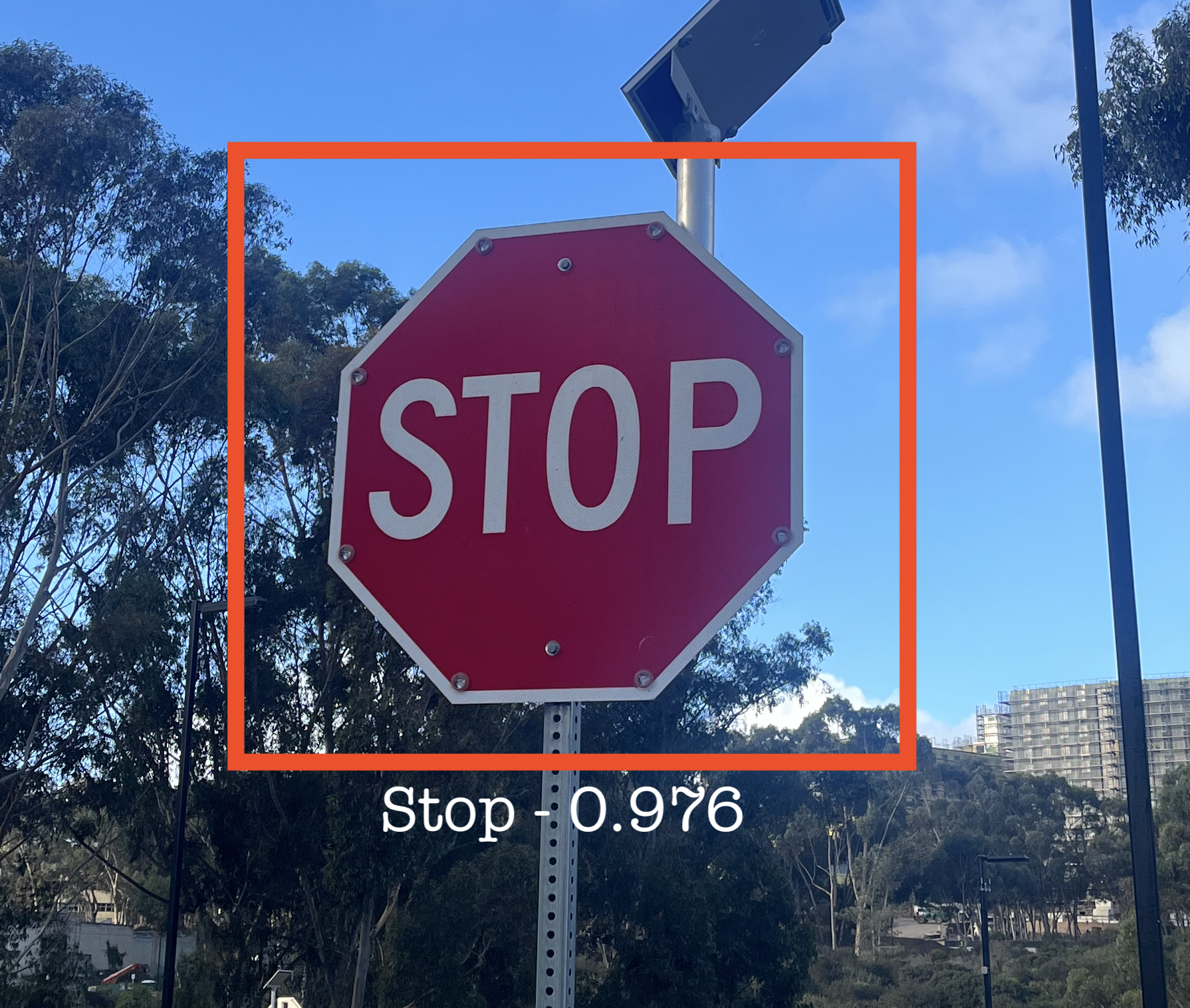}
        \subcaption{Clean Image}
        \label{fig:stopsign1}
    \end{minipage}
    \hfill
    \begin{minipage}{0.45\linewidth}
        \centering
        \includegraphics[width=\linewidth]{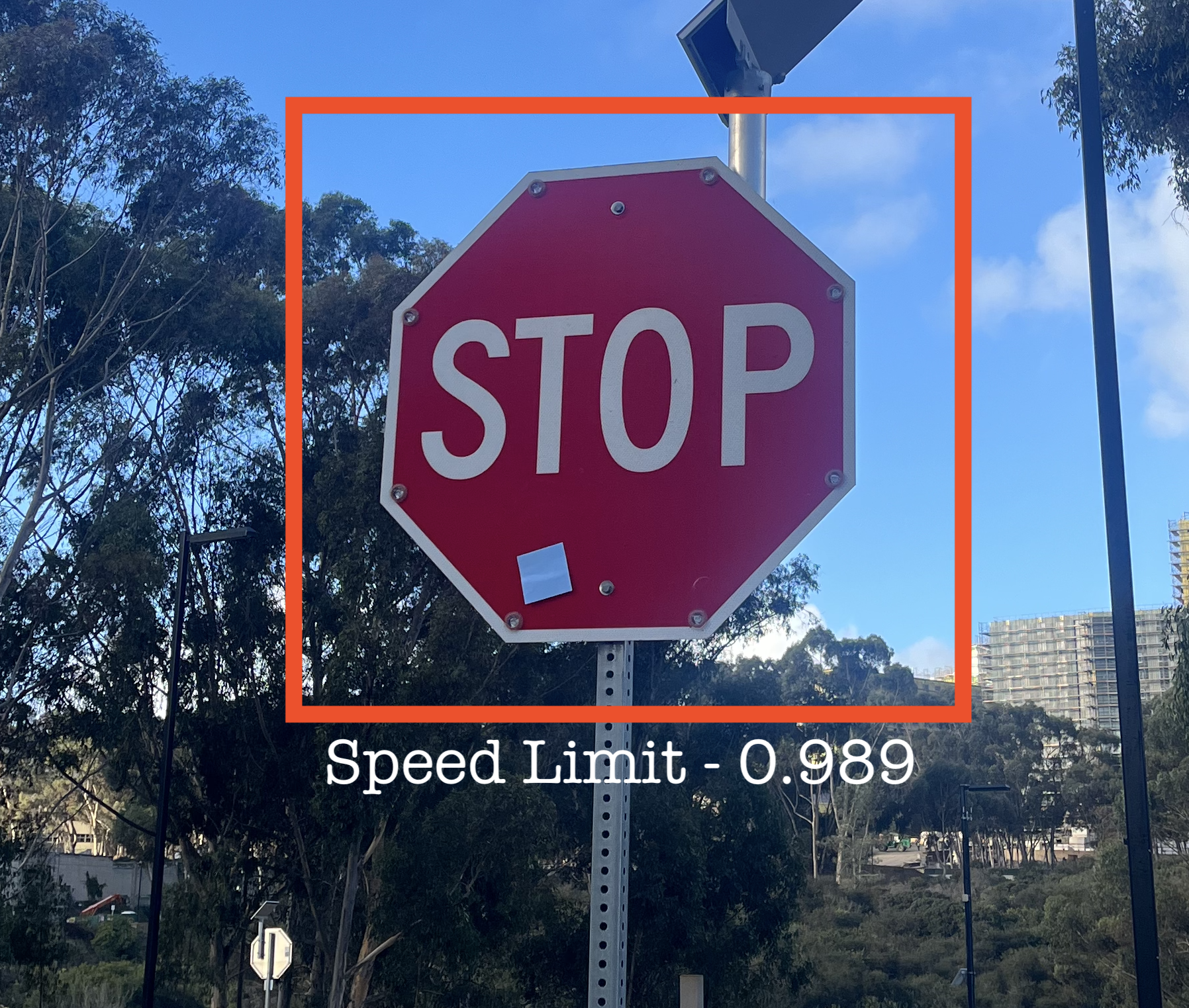}
        \subcaption{Trigger Embedded Image}
        \label{fig:stopsign2}
    \end{minipage}
    \caption{A machine learning model may be trained to wrongly classify images with triggers such as post-it notes to speed limit signs. This could lead to dangerous misclassifications of stop signs, potentially preventing a car from stopping if used in an autonomous vehicle.}
    \label{fig:stopsigns}
\end{figure}

Trojan attacks - also referred to as \emph{backdoor attacks} - have emerged as a tangible and critical threat to the safe use of deep neural networks in the real world. While adversarial attacks refer to those which perturb images at inference time \cite{perturb}, trojan attacks involve training a model to misclassify inputs embedded with triggers to an attacker's target class. These attacks have become especially concerning in higher stakes scenarios such as in medical diagnostics or self driving systems\cite{autoDriving, medical}. A classic scenario of trojaning, as discussed in the foundational paper \textit{BadNets} \cite{badnets}, is the misclassification of a stop sign to a speed limit sign. Upon deployment, simply placing a trigger item in view can cause a trojaned model to behave maliciously. This example is shown in \textit{Figure} \ref{fig:stopsigns} and demonstrates real-world vulnerabilities. Even discrete and inconspicuous modifications to physical objects in images can exploit trojans within machine learning models, leading to misclassifications with severe consequences.

\subsection{Types of Attacks}

Gu et al. \cite{badnets} introduce one of the earliest backdoor attacks in deep learning, \textit{BadNets}. This attack involves the \textit{poisoning} of a dataset. Given a training dataset $\mathcal{D} = \{(\mathbf{x}_i, y_i)\}$, a subset of samples is selected and a trigger pattern $\mathbf{t}$ is embedded into these images to create poisoned samples $\mathbf{x}_i' = \mathbf{x}_i + \mathbf{t}$. These modified samples are then relabeled with the attacker-specified target label $y_t$. The poisoned dataset $\mathcal{D}' = \{(\mathbf{x}_i', y_t)\} \cup \{(\mathbf{x}_i, y_i)\}$ is then used for training the trojaned model. The resulting model performs well on benign samples with accuracy similar to an untrojaned model but misclassifies any input containing the trigger pattern $\mathbf{t}$ as $y_t$. BadNets was a salient work which demonstrated the feasibility of visible backdoor attacks and laid the groundwork for many subsequent poisoning-based attacks, illuminating the critical security risks posed by these methods.

Chen et al \cite{blended} assume a stricter threat model in which the attacker has no knowledge of the model or training data but can add a small number of samples to the model's training set. The authors propose \textit{Blended}, an attack where an adversary injects a small number of poisoned samples into the training data to cause a model to misclassify poisoned inputs to a target class. The authors demonstrate the feasibility of their method, using as few as 50 poisoned samples to effectively trojan a model.

Liu et al \cite{trojannn} propose an alternative threat model in which an attacker does not have access or the ability to add to the training data, but is able to retrain the model. Their attack, \textit{TrojanNN}, involves inverting the neural network to generate a general trigger and then retraining the model on reversed engineered training data to establish a causal link between the trigger and the desired malicious output. The authors demonstrate the efficacy of their attack method on various applications such as face recognition and autonomous driving.

While previously mentioned attacks involve the use of human-detectable triggers, Nguyen et al \cite{wanet}  propose the attack \textit{WaNet}, which leverages image warping to create less perceptible triggers. Unlike previous backdoor attacks that rely on more visible perturbations such as patches or blending, WaNet uses a small and smooth warping field to deform the input image in a way that is hard for humans to detect. The authors show that their warping-based backdoor outperforms previous methods in a human inspection test, demonstrating its stealthy nature. The resulting trojaned models are effective at attacking and bypassing state-of-the-art defenses on standard classification datasets.

In the work by Barni et al \cite{sig}, the authors present signal backdoor attack (SIG). SIG is free of label poisoning and relies solely on perturbations within the input features. The motivation of this is to prevent trojan detection via visual inspection, as label poisoning could easily be detected by auditing the data prior to training.

\subsection{Mitigation Strategies}

A naive but straightforward method for Trojan mitigation is fine-tuning. In this approach, a user retrains a potentially compromised model using a smaller, clean dataset that is available to them. Fine-tuning aims to adjust the model’s weights, potentially overriding the malicious influence of the backdoor trigger. Liu et al. \cite{neuraltrojans} suggest that this method can often reduce the risk associated with trojaned models by lessening the impact of the malicious weights embedded within the model. Contrary to retraining scratch, fine-tuning typically requires fewer training samples and converges more quickly, making it a computationally efficient alternative in comparison to full retraining. However, the success of this approach depends on the size and quality of the clean dataset used for fine-tuning, and potentially the nature of the trojan attack and model architecture.

Zhu et al \cite{ftsam} propose \textit{FTSAM}, which leverages sharpness-aware minimization (SAM) to more effectively fine-tune a trojaned model with the goal of removing the model’s backdoors. The authors find that vanilla fine-tuning fails to significantly change the weight norms of backdoor-related neurons, which are key to the backdoor behavior. The authors claim that incorporating SAM into the retraining process effectively perturbs the backdoor-related neurons, allowing the model to escape the local minimum and better achieve mitigation.

Wang et al introduce \textit{Neural Cleanse} \cite{neuralcleanse}, which includes a two-part mitigation system with strong generalization. The first component is a filter that identifies adversarial inputs based on their neuron activation patterns. This filter utilizes the neuron activation profile of each label to reverse-engineer a trigger, setting a threshold to distinguish clean inputs from adversarial ones. The second component involves patching the compromised model using a technique called neuron pruning. This method targets neurons that exhibit significant activation differences when processing clean versus adversarial inputs. By pruning these neurons, the method reduces the model's responsiveness to backdoor triggers while minimizing any impact on classification accuracy. While Neural Cleanse is an effective mitigation strategy, the first stage is computationally prohibitive for models with large number of classes.

Liu et al \cite{fineprune} propose a mitigation tactic combining fine-tuning and pruning called \textit{Fine-Pruning}. First all the activations are recorded from running the benign data, based on these activations some neurons are targeted for pruning. Pruning is first applied to remove the backdoor-related neurons. Fine-tuning is then applied in order to restore the lost model accuracy. The iterative pruning makes this method computationally expensive.

\section{Methodology}

\subsection{Threat Model}

\subsubsection{Attacker}
We assume the presence of an adversary conducting a backdoor attack on a deep neural network (DNN). In this context, the \textit{poisoning ratio} represents the fraction of compromised samples present in the training dataset. The attacker's objective is to manipulate the model such that it consistently classifies inputs containing specific triggers as target labels while preserving correct classification behavior on unaltered, clean samples.

\subsubsection{Defender’s Objective}
The defender obtains a potentially backdoored model and is unable to train their own model from scratch given constrained resources, either in terms of computation or data requirements. We assume the defender has access a small set of benign data samples, denoted as $D_{benign}$. The defender’s goal is to alter the model in such a way that, should it be compromised, it retains high performance on the benign dataset while effectively eliminating any potential backdoors within the model. Specifically, this means minimizing the proportion of poisoned inputs misclassified as the target label.

\subsection{Key Intuition}
The process of fine-tuning a trojaned model using a limited number of clean data samples can cause minor adjustments to model weights, thereby reducing the attack success rate (ASR) while preserving the model's accuracy on benign data. Previous work by Zhu et al.~\cite{ftsam} suggests that standard fine-tuning is insufficient for removing backdoors, as the backdoored model already fits benign training data, causing weight updates through gradient descent to be minimal and insufficient in mitigating the attack.

Building on these findings, we propose a regularization-based post training compression technique that targets specific layers within the model. This approach perturbs the weights strategically during fine-tuning, ensuring substantial weight adjustments across all layers. Our method subsequently restores the model's performance post-compression, resulting in a compressed, clean model that maintains high accuracy. We elaborate on the details of our proposed approach, \sys, in the subsequent sections.

\subsection{\sys{}}

\begin{figure*}[ht]
    \centering
    \includegraphics[width=\textwidth]{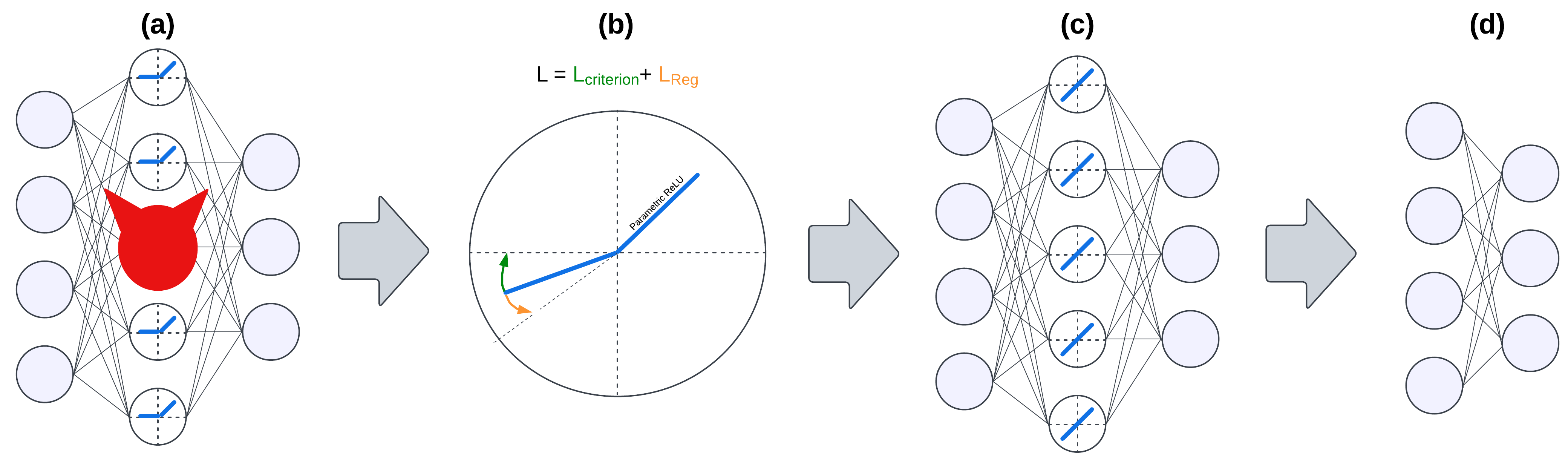}
    \caption{Illustration of the proposed \sys{} methodology for mitigating trojans in neural networks. (a) Identification of a potentially compromised layer suspected to contain a trojan. (b) Application of fine-tuning with the regularization strategy described in Section~\ref{subsec:method}, designed to incrementally guide the activation towards linearity. (c) Post-tuning, the activation function approximates an identity function. (d) Simplification of the network architecture by merging two consecutive fully connected layers into a single, reduced layer, effectively eliminating the intermediary layer.}

    \label{fig:methodology}
\end{figure*}

\label{subsec:method}
In this section, we introduce \sys, a novel regularization strategy aimed at adaptively removing layers from fully connected and convolutional layers in deep neural networks. By removing layers, this method compresses the model and effectively mitigates backdoor attacks in the process. Unlike conventional approaches that may rely heavily upon targeting backdoor-related neurons in the network, \sys{} focuses on removing the layers that are prone to being trojaned, cleansing the model in the process. This method changes the model weights through a regularization technique, making sufficient changes to the weights of the model removing the effect of triggers and backdoor attacks.


As demonstrated in Figure~\ref{fig:methodology}, \sys{} involves regularizing the complexity of individual layers by linearizing their activation functions during training or fine-tuning. By converting these activation functions to a linear form, we can merge consecutive linear layers into a single equivalent layer, thereby reducing the model's depth. This results in a model with lower storage and computational requirements, independent of specialized software or hardware optimizations.

The regularizer is designed to adaptively modulate the negative slope of \textit{multilayer perceptrons} (MLPs) or \textit{convolutional} layers. When the negative slope of a given layer reaches one, we can safely merge the two consecutive linear transformation into one linear transformation effectively eliminating one layer from the network without any loss in performance. This is due to the fact that two consecutive linear transformations with no \textit{non-linearity} between them are equivalent to a one linear transformation that can be represented using one linear layer. This yields a shallower network with fewer parameters, subsequently reducing resource demands during execution. We demonstrate that this transformation significantly alters the model state, effectively neutralizing backdoor effects and producing a clean, high-accuracy, and computationally efficient model.

\subsection{Formal Definition}
\label{subsec:formal_def}
We will first derive all results for fully connected layers. We then will extend the result to Convolutional layers.
We define a \textit{mergeable layer} as two consecutive linear layers that use a Parametric Rectified Linear Unit (PReLU) \cite{he2015delving} as their activation function. The PReLU function is expressed as follows:
\begin{equation}
\label{eq:relu}
PReLU_\alpha(x) = \max(0, x) + \alpha \min(0, x),
\end{equation}
where $\alpha$ denotes the slope of the PReLU function when $x$ is negative.

The output of a basic two-layer mergeable layer can be described as:
\begin{equation}
Y_\alpha = W_2(PReLU_\alpha(W_1X + b_1)) + b_2,
\end{equation}
where $X$ is the input random variable, $Y$ is the output random variable, $W_1$ and $W_2$ represent the weight matrices, and $b_1$ and $b_2$ are the bias vectors for the first and second fully connected layers, respectively. By adjusting $\alpha$, the activation function can be made linear.

We define $Y_{linear}$ as:
\begin{equation}
Y_{linear} = W_2W_1X + W_2b_1 + b_2.
\end{equation}
We define the \textit{non-linearity error} as the squared difference between $Y_\alpha$ and $Y_{linear}$.

It can be shown that for any $\delta \in [0,1]$ and for any distribution of input random variable $X$:
\begin{equation}
\label{eq:p_error}
P\left\{ |Y_{\text{linear}} - Y_\alpha|^2 \leq C \times (1 - \alpha)^2 \right\} > 1 - \delta,
\end{equation}
where
\begin{equation}
\begin{aligned}
&x^\delta \equiv \inf\left\{x \in \mathcal{R}^n \mid P\left\{ |X| > |x| \right\} < \delta\right\}, \\
&C = \sigma_{\max}(W_2W_1)^2 |x^\delta|^2 + |W_2b_1|^2,
\end{aligned}
\end{equation}
and $\sigma_{\max}(W_2W_1)$ is the largest singular value of $W_2W_1$. By choosing $\delta$ arbitrarily small, the probability of error is bounded proportional to $(1 - \alpha)^2$ with probability one.

To optimize for low non-linearity error We incorporate the term $(1 - \alpha)^2$ into our loss function as a regularizer to jointly optimize for both non-linearity reduction and the label cross-entropy loss. This loss term aligns $Y_\alpha$ with $Y_{linear}$, thereby compressing the model into a shallower form. The compressed layer's weights and bias can be calculated using the formula for calculating $Y_{linear}$.

\begin{table*}[htbp] 
\centering
\caption{Performance of Various Attack/Cleansing Methods on PreAct-ResNet18}
\label{tab:preact}
\begin{tabular}{lcc|cccccccccc}
\toprule
\textbf{Attack/Cleansing} & \multicolumn{2}{c}{\textbf{Trojaned}} & \multicolumn{2}{c}{\textbf{FT}} & \multicolumn{2}{c}{\textbf{FP}} & \multicolumn{2}{c}{\textbf{FTSAM}} & \multicolumn{2}{c}{\textbf{NC}} & \multicolumn{2}{c}{\textbf{MG}} \\
\cmidrule(r){2-3} \cmidrule(l){4-5} \cmidrule(l){6-7} \cmidrule(l){8-9} \cmidrule(l){10-11} \cmidrule(l){12-13}
 & \textbf{Test Acc} & \textbf{ASR} & \textbf{Test Acc} & \textbf{ASR} & \textbf{Test Acc} & \textbf{ASR} & \textbf{Test Acc} & \textbf{ASR} & \textbf{Test Acc} & \textbf{ASR} & \textbf{Test Acc} & \textbf{ASR} \\
\midrule
TrojanNN & 90.90 & 99.98 & 88.23 & 15.16 & \textbf{91.86} & 98.44 & 91.47 & 1.38 & 90.44 & \textbf{1.17} & 89.09 & 3.88 \\
\cmidrule(lr){1-13}
WaNet & 87.38 & 97.41 & 87.94 & 6.34 & \textbf{91.44} & 6.73 & 91.41 & 4.08 & 87.38 & 97.41 & 90.20 & \textbf{1.53} \\
\cmidrule(lr){1-13}
BadNet & 90.44 & 94.03 & 85.59 & 1.01 & 91.90 & 3.77 & 91.25 & 1.67 & 91.45 & \textbf{0.87} & \textbf{93.50} & 11.37 \\
\cmidrule(lr){1-13}
Blended & 91.09 & 99.67 & 88.03 & 8.74 & \textbf{92.08} & 58.51 & 90.97 & 9.76 & 89.53 & \textbf{2.08} & 88.20 & 3.73 \\
\cmidrule(lr){1-13}
SIG & 82.52 & 97.49 & 88.01 & 5.37 & 89.63 & 1.34 & \textbf{90.14} & \textbf{0.06} & 82.52 & 97.49 & 88.34 & 0.14 \\
\cmidrule(lr){1-13}
\textbf{Average} & \textbf{88.47} & \textbf{97.72} & \textbf{87.56} & \textbf{7.32} & \underline{\textbf{91.38}} & \textbf{33.76} & \textbf{91.05} & \underline{\textbf{3.39}} & \textbf{88.26} & \textbf{39.80} & \textbf{89.87} & \textbf{4.13} \\

\bottomrule
\end{tabular}
\end{table*}

\begin{table*}[htbp] 
\centering
\caption{Performance of Various Attack/Cleansing Methods on Vision Transformer (ViT)}
\label{tab:vit}
\begin{tabular}{lcc|cccccccccc}
\toprule
\textbf{Attack/Cleansing} & \multicolumn{2}{c}{\textbf{Trojaned}} & \multicolumn{2}{c}{\textbf{FT}} & \multicolumn{2}{c}{\textbf{FP}} & \multicolumn{2}{c}{\textbf{FTSAM}} & \multicolumn{2}{c}{\textbf{NC}} & \multicolumn{2}{c}{\textbf{MG}} \\
\cmidrule(r){2-3} \cmidrule(l){4-5} \cmidrule(l){6-7} \cmidrule(l){8-9} \cmidrule(l){10-11} \cmidrule(l){12-13}
 & \textbf{Test Acc} & \textbf{ASR} & \textbf{Test Acc} & \textbf{ASR} & \textbf{Test Acc} & \textbf{ASR} & \textbf{Test Acc} & \textbf{ASR} & \textbf{Test Acc} & \textbf{ASR} & \textbf{Test Acc} & \textbf{ASR} \\
\midrule
TrojanNN & \textbf{98.03} & 100.0 & 95.98 & 100.0 & 96.90 & 100.0 & 97.33 & 100.0 & 97.05 & \textbf{0.72} & 92.70 & 7.39 \\
\cmidrule(lr){1-13}
WaNet & 94.45 & 84.73 & 94.56 & 0.71 & \textbf{97.52} & 0.26 & 94.11 & 27.69 & 97.11 & \textbf{0.18} & 94.13 & 1.62 \\
\cmidrule(lr){1-13}
BadNet & 95.85 & 93.44 & 94.11 & 27.69 & 97.08 & 65.53 & \textbf{97.20} & 89.07 & 96.76 & \textbf{0.27} & 93.55 & 11.38 \\
\cmidrule(lr){1-13}
Blended & \textbf{98.16} & 99.96 & 94.39 & 17.00 & 31.56 & 5.31 & 98.09 & 99.70 & 96.93 & 47.00 & 92.71 & \textbf{3.67} \\
\cmidrule(lr){1-13}
SIG & 86.82 & 89.44 & 94.61 & 35.93 & \textbf{95.96} & 87.89 & 95.02 & 93.12 & 96.31 & 81.01 & 92.94 & \textbf{1.99} \\
\cmidrule(lr){1-13}
\textbf{Average} & \textbf{94.66} & \textbf{93.52} & \textbf{94.73} & \textbf{36.27} & \textbf{83.80} & \textbf{51.80} & \textbf{96.35} & \textbf{81.92} & \underline{\textbf{96.83}} & \textbf{25.84} & \textbf{93.21} & \underline{\textbf{5.21}} \\

\bottomrule
\end{tabular}
\end{table*}

\subsection{Compression Ratio}
The compression ratio of a reduced fully connected layer is expressed as:
\begin{equation}
\label{eq:compression_ratio}
CR = 1 - \frac{n_{in} \times n_{out}}{n_{hidden} \times (n_{in} + n_{out})}
\end{equation}
where $CR$ denotes the compression ratio, with $n_{in}$, $n_{hidden}$, and $n_{out}$ representing the input size, hidden size, and output size of a 2-layer MLP, respectively.

Note that the compression ratio may be negative, indicating an expansion rather than a compression. A positive compression ratio is achieved only if $n_{hidden}$ exceeds the ratio of the product to the sum of $n_{in}$ and $n_{out}$. This highlights that MLPs with a ``bottleneck" configuration (small $n_{hidden}$) may not benefit from this method. Conversely, MLPs with a ``widening" structure typically exhibit favorable compression ratios.

\subsection{Convolutional Layers}
Convolutional layers, which are critical to many neural architectures, can be treated as a fully connected layer with weight redundancies. Given their linear transformation properties, they can be represented using circulant matrices, where each row is a cyclic permutation of its predecessor. Merging two convolutional layers results in a single convolution layer where the kernel size, $k$, is given by:
\begin{equation}
k = k_1 + k_2 - 1
\end{equation}
where $k_1$ and $k_2$ are the kernel sizes of the original layers.

The compression ratio for convolutional layers can be determined as:
\begin{equation}
CR_{conv} = 1 - \frac{k_1^2 \times c_{in} \times c_{hidden} + k_2^2 \times c_{hidden} \times c_{out}}{(k_1 + k_2 - 1)^2 \times c_{in} \times c_{out}}
\end{equation}
where $c_{in}$, $c_{hidden}$, and $c_{out}$ represent the input channels, hidden channels, and output channels, respectively. However, a pivotal consideration for common CNN architectures like ResNet is the potential for a negative compression ratio, emphasizing the need for careful considerations when using \sys{} for convolutional networks. 
However, we show the effectiveness of \sys{} as an effective backdoor mitigation technique in section ~\ref{sec:eval}.

\subsection{Alternative Activation Functions}
Our discussion has primarily centered around the ReLU activation function, known for its simplicity and effectiveness. Recent advances have seen the emergence of ReLU variants like ELU \cite{clevert2015fast}, GeLU \cite{hendrycks2016gaussian}, and SiLU \cite{elfwing2018sigmoid}, which offer enhanced adaptability.

We introduce a method to linearize ReLU-family activations. For instance, in the ELU function, which is linear in the positive domain and non-linear in the negative, we integrate a linear component, $y = x$, into the negative domain. This adjustment is formulated as a weighted combination:
\[
\text{ELU}^\beta(x) = 
\begin{cases} 
      x & \text{if } x > 0 \\
      \alpha x + (1 - \alpha) \beta (\exp(x) - 1) & \text{if } x \leq 0 
\end{cases}
\]
where $\beta$ represents the parameter of the ELU function. The weight $\alpha$, constrained to [0, 1] via a sigmoid function during training, dictates the balance between linear and non-linear components.

For one-dimensional activation functions like GeLU and SiLU, characterized by the form $f(x) = x \times h(x)$, where $h(x)$ transitions between bounds as $x$ varies, we modify these functions using $\alpha$ to yield parametric forms:
\[
\text{GeLU}^\alpha(x) = x(\Phi(x) + \alpha (1 - \Phi(x)))
\]
with $\Phi(x)$ being the Gaussian cumulative distribution function, and for SiLU:
\[
\text{SiLU}^\alpha(x) = x(\sigma(x) + \alpha (1 - \sigma(x)))
\]
where $\sigma(x)$ denotes the sigmoid function.

\section{Evaluation}
\label{sec:eval}

\subsection{Experimental Setup}

\textbf{Attack Configuration.} We investigate the efficacy of our method against five leading trojan attacks: TrojanNN \cite{trojannn}, WaNet \cite{wanet}, BadNet \cite{badnets}, Blended \cite{blended}, and SIG \cite{sig}. These methods were chosen for their demonstrated proficiency in evading conventional security mechanisms while preserving high testing accuracy. Experiments are conducted using the CIFAR10 dataset \cite{cifar10} with predefined parameters from BackdoorBench \cite{backdoorbench} to maintain consistency. We poison the models at a 10\% rate, targeting the 0th label. Our tests include two representative architectures from the major families of vision models: PreAct-ResNet18 for CNNs and ViT-base-16 for vision transformers.

\textbf{Defense Configuration.} We test \sys{} against standard fine-tuning (FT) and three leading backdoor mitigation methods: Fine-Pruning (FP) \cite{fineprune}, FTSAM \cite{ftsam}, and Neural Cleanse (NC) \cite{neuralcleanse}. For these experiments, 5\% of the clean dataset is utilized. Training is conducted using the SGD optimizer with a momentum of $0.9$, spanning over $20$ epochs on clean data with a batch size of 128. In the spirit of fair comparison, we conducted grid search for each learning rate parameter, selecting the one that achieved the greatest reduction in ASR for that method. The exhaustive configurations of each experiment are given in our code base. Each defense method begins from the same compromised baseline model. In our approach, presuming the trojan's presence in the terminal layers based on the conjectures of previous works \cite{trojan3}, we apply regularization to the last three layers of the PreAct-Resnet18 and the last four layers of the Vision Transformer. The regularization strength used in all experiments is one. Removing 4 fully connected layers from the base ViT variant will reduce the parameter count from $85.8$ million parameters down to $73.4$ Million parameters resulting in a $15\%$ reduction in model size.

\textbf{Defense Metric.} The efficacy of our trojan mitigation strategy is measured by means of two metrics: model test accuracy (Test Acc) and Attack Success Rate (ASR). The ASR gauges the potency of a trojan in forcing the model to misclassify a designated label. An effective mitigation strategy is reflected by a reduced ASR. The goal is the preserve the Test Accuracy while reducing ASR.
Another metric we consider is the computational cost of running each of these experiments. Some of the defenses, namely Neural Cleanse and Fine-Pruning, have a very expensive pre-processing stage that, depending on the application, may be impractical. The average runtime cost of each method is given in figure \ref{fig:computeTime}.





\subsection{Evaluation of Results}

\subsubsection{Efficacy of Trojan Mitigation in Vision Transformers}
The robustness of ViT to various trojan mitigation strategies was examined and a comparison to the robustness of CNNs is shown between Tables \ref{tab:preact} and \ref{tab:vit}. The trojaned ViT models demonstrated significant resilience against most existing cleansing techniques. In contrast, PreAct-ResNet18 responded more favorably to these methods. However, \sys{} exhibits a model-agnostic property which is extends to ViT models. As detailed in Table \ref{tab:vit}, while other methods sporadically mitigated trojans in ViT depending on the attack method used, \sys{} consistently outperformed these techniques in terms of consistent efficacy across attacks.

\subsubsection{Computational Efficiency of \sys{}}
\begin{figure}[tbp]
    \centering
    \includegraphics[width=\linewidth]{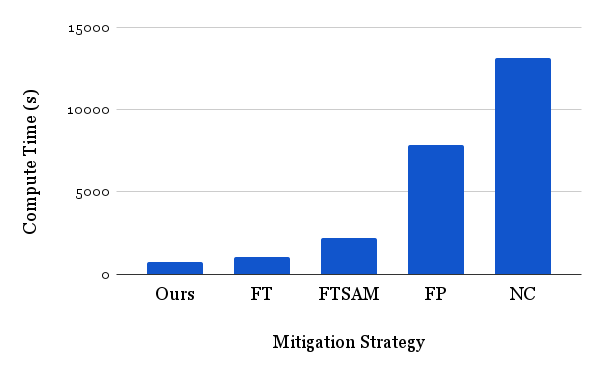} 
    \caption{Computational time for mitigation strategies on ViT.} 
    \label{fig:computeTime} 
\end{figure}

The computational efficiency of \sys{} and other mitigation strategies was evaluated based on the processing time required for each method. In all experiments, a single A6000 GPU was used. \textit{Figure} \ref{fig:computeTime} portrays the processing time taken for each method. \sys{} displayed superior performance, achieving computational speedups of $10.6x$, $17.7x$, and $3.0x$ compared to Fine-Pruning (FP), Neural Cleanse (NC), and Feature Squeezing and Model Augmentation (FTSAM) respectively. While FP and NC showed efficacy against many of the attacks tested, their computational demands were substantially higher, highlighting the efficiency of \sys{} as a preferable choice for practical applications. Additionally, the resulting cleansed model produced by \sys{} has $73.4$ million parameters, $15\%$ less than the original model with $85.8$ million parameters. This reduction in parameter count is accompanied by a $14\%$ drop in the number of MAC operations.



\section{Conclusion}

In this work we conducted experiments demonstrating the robustness of trojaned Vision Transformers to SOTA cleansing methods. We showed these methods are far less reliably effective on transformer based architectures for different attacks. We introduced \sys{}, an innovative compression based approach for countering AI Trojan attacks. \sys{} provides an architecture-agnostic trojan mitigation strategy applicable to, not just CNNs, but also Transformer models for which \sys{} reduced the attack success rate by more than $20\%$ across existing state-of-the-art. 


Our analysis of runtime efficiency demonstrates that \sys{} requires significantly less compute time than other defense strategies and additionally doubles as an effective compression method, achieving parameter reductions up to 15\% and MAC reductions up to $14\%$ with negligible impact on accuracy.

The open-source code for \sys{} is made publicly accessible\footnote{https://github.com/yjandali/BackdoorBench-MergeGuard}. 
Future research directions include developing algorithms for the selection of \sys{} layers, exploring alternative regularization techniques, and extending the method to additional transformer-based architectures.

\bibliography{example_bib}
\bibliographystyle{ieeetr}

\end{document}